\documentclass{ws-ijmpa}
\begin{document}
\markboth{Nilo Sylvio Costa Serpa}
{}
%
%
\title{MODELLING THE DYNAMICS OF THE WORK-EMPLOYMENT SYSTEM BY PREDATOR-PREY INTERACTIONS}
\author{N. S. COSTA SERPA}
\address{UNIP - Universidade Paulista, Instituto de Ci\^{e}ncias Exatas e Tecnologia\\
SGAS Quadra 913, s/n$^o$ - Conjunto B - Asa Sul - Bras\'{i}lia - DF, Brasil,
CEP 70390-130\\
nilo.serpa@mte.gov.br, \
nilo@techsolarium.com}
\author{J. R. STEINER}
\address{UnB - Universidade de Bras\'{i}lia, Instituto de F\'{i}sica\\
CP 04455, Bras\'{i}lia - DF, Brasil,
CEP 70919-970\\
jrsteiner78@gmail.com}
\maketitle
\begin{abstract}
The broad application range of the predator-prey modelling enabled us to apply it to represent the dynamics of the work-employment system. For the adopted period, we conclude that this dynamics is chaotic in the beginning of the time series and tends to less perturbed states, as time goes by, due to public policies and hidden intrinsic system features. Basic Lotka-Volterra approach was revised and adapted to the reality of the study. The final aim is to provide managers with generalized theoretical elements that allow to a more accurate understanding of the behavior of the work-employment system.
\end{abstract}
\keywords{predator-prey model; chaotic dynamics; work-employment system; perturbed states; evolution.}
\paragraph\
\paragraph\
\begin{flushright} 
\footnotesize {"Economies possess general ecosystem properties, such as dynamism, evolution, integrity, stability and resilience. Economies are inextricably embedded in larger natural ecosystems, and exchange flows of materials and energy with natural systems".}
\end{flushright}
\paragraph\
\begin{flushright} 
\footnotesize {S. Farber \& D. Bradley}
\end{flushright}
\section{Introduction} 
The concept of equilibrium in the predator-prey population dynamics has its origins in the works of Lotka \cite{Lotka} and Volterra \cite{Volterra}, making the base of several theoretical models of interaction among species, with applications in the context of ecological systems and wildlife management (Caughley and Sinclair, 1994 \cite{Caughley}; Earn {\it et al.}, 2000 \cite{Earn}; Blasius {\it et al.}, 1999 \cite{Blasius}), including fluctuating environments (Collie and Spencer, 1997 \cite{Spencer}). Particularly, Blasius {\it et al.} were very benefited by previous works on phase synchronization phenomena in coupled chaotic systems (Rosenblum et al., 1996 \cite{Rosenblum}). Also, there are generalized approaches of the
Lotka-Volterra model (Tu and Wilman, 1992 \cite{Tu}). There is special interest on the problem of the growth of two populations conflicting with one another, known as "the problem of Volterra" \cite{Davis}. 
The generality of type predator-prey models becomes possible to abstract them from their early ecological roots, and, by analogy, to apply them on a large range of mathematical modelling problems. Recently it befell to us to investigate the work-employment dynamics in Brazil by the implementation of a type predator-prey model. It is not the case of a classical application of the standard Lotka-Volterra competition modelling, since the variables storage dissimilarities among population registers, not the populations themselves; this is so because the existing cyclic pattern which entangles the work-employment evolution is much more easily discerned by dissimilar modelling. Besides, while is tempting to discard the standard Lotka-Volterra model as too simplistic, there is a real situation where the detailed and complex dissimilar modelling holds undeniable utility.
The aim of this article is to show that the work-employment system in Brazil admits a predator-prey modelling. The model started from a simple correlation suggested by the superposition of two time series, one to the number of employed workers, and other to the number of active employers. The entanglement of the two series is analyzed with the aid of the clustering methodology. In the present article, the abstract analogy with a natural ecosystem is the work market; the employer or corporation plays the rule of the predator; the prey is represented by the worker. 
The assumption of employers as predators has many reasons:
\begin{itemize}
\item The number of employers is much smaller than the number of workers (as in linx-hare, lion-zebra, crocodile-gnu and cheetah-baboon predator-prey relationships).
\item The employers handle the number of workers depending on the profit margins which they want, heavily controlling the levels of unemployment.
\item Employers often fail to collect the so-called "Guarantee Fund for Time in Service" related to their workers, causing irregularities that hinder the realization of labor rights in the event of contract termination agreement. Moreover, even without a contract being rescinded, the nonpayment of the Guarantee Fund blocks the employee to request it to buy a home.
\item Employers often fail to inform the CAGED (Cadastro Geral de Empregados e Desempregados)\footnote{CAGED is a great database updated monthly and containing all workers and employers in Brazil.} on the movement of their staff, resulting in bureaucratic disarray that blocks the access to some labor rights in the event of loss of working papers; furthermore, the fail to inform the CAGED affects the labor statistics.
\item Many employers absorb the workforce without formal contracts, depriving workers of their entitlements.
\item However the scenario is changing in Brazil, many employers still do not make investments in quality, training employees and providing opportunities for better wages by productivity and capacity. The result is a high turn-over, low wages and a working class that never achieves a status of participation on the employer's profits. In sum, we have a highly non-egalitarian society which does not prioritize education and manpower quality. Socially speaking, the predator is who foment social discrepancies and, in present case, they are employers (public or private). That is the point here.
\end{itemize}
From this point of view, we have not a model based on political preferences, as some watchers would think, but based on hard facts. There is no influence of maniqueist ideas, but simply findings. Paraphrasing Lawson, the offer of labor posts depends on the investment decisions of capitalists (employers), and the investment decisions of capitalists depend on the existing offer of manpower \cite{Lawson}. Thus, since employers impose financial directions on the work market, often indifferent to the zeal for the rights of the working class, thus creating distortions and fueling widespread dissatisfaction, it is reasonable to consider the adopted viewpoint in constructing the model. Obviously, in a more egalitarian society, firmly established on ethics and education, economically viable and with eyes in the future, it would be required a comprehensive review of the basic premises assumed. Also, in face of the facts itemized above, employers appear in many situations to dampen fluctuations in employee populations.
\section{Observational premises} 
The data sources were the CAGED, and the RAIS (Rela\c{c}\~{a}o Anual de Informa\c{c}\~{o}es Sociais), two of the more important corporative databases of the Brazilian Ministry of Work and Employment. First of all we looked for harmonic patterns in the behavior of the working class and active employers. Figure 1 clearly presents the referred pattern in the chosen period (1996-2008). We defined active employer or company as one that presents variation in the number of admitted workers from a certain month to another. Secondly, we verify the presence of lags between the harmonic patterns of both series. This feature is also clearly visible in figure 1, with relaxation of the system all along the last fifty months. Table 1 shows a segment of the CAGED for the first year of the time series considered in present study
\begin{center}
\begin{figure} [h]
\includegraphics[scale=0.25]{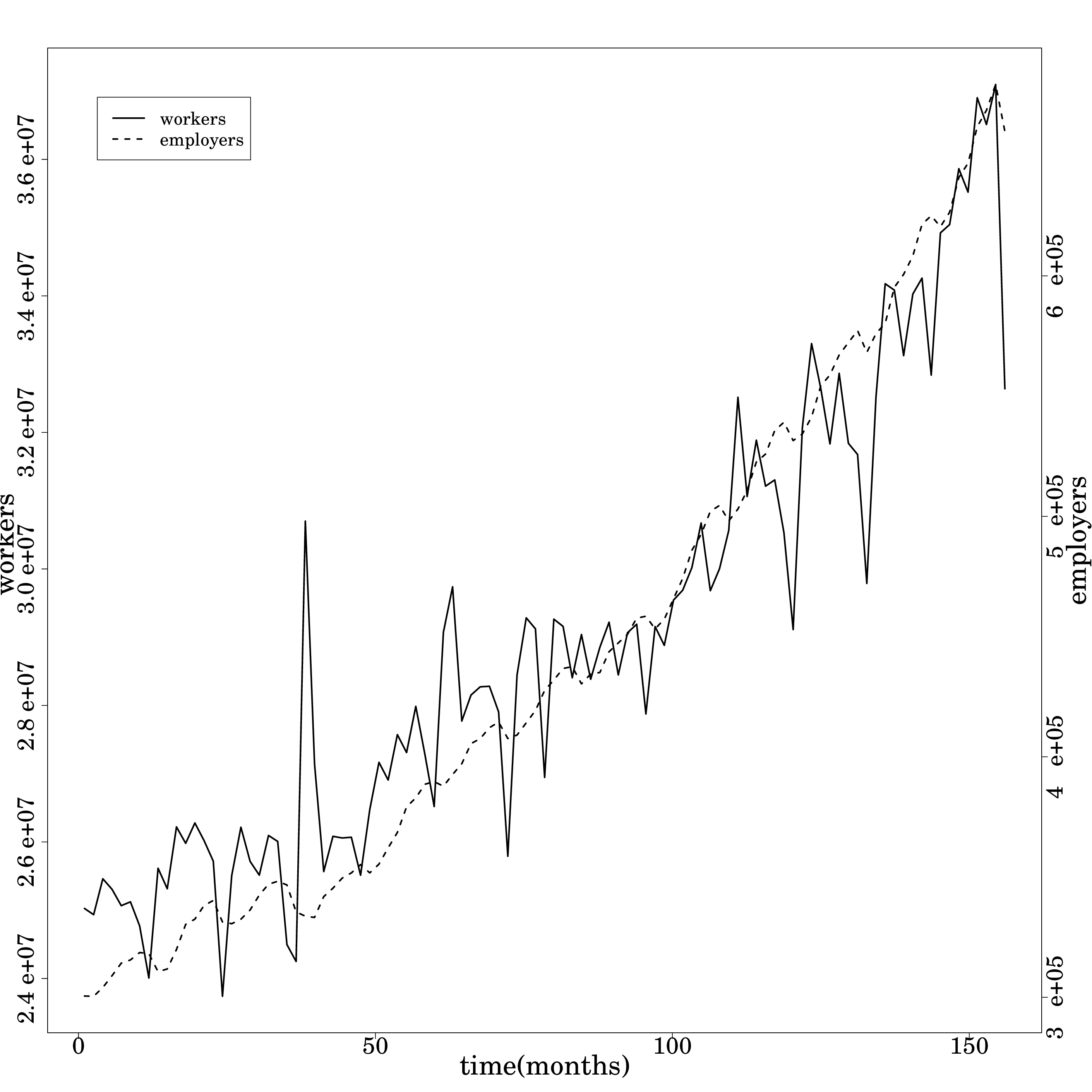}
\\
\small {Figure 1: {\it harmonic patterns from workers (full line) and active employers (dashed line). The phase difference is clear in most of the peaks and valleys, especially in the fifty more recent months.}} 
\end{figure}
\end{center}
\begin{figure} [h]
\includegraphics[scale=0.48]{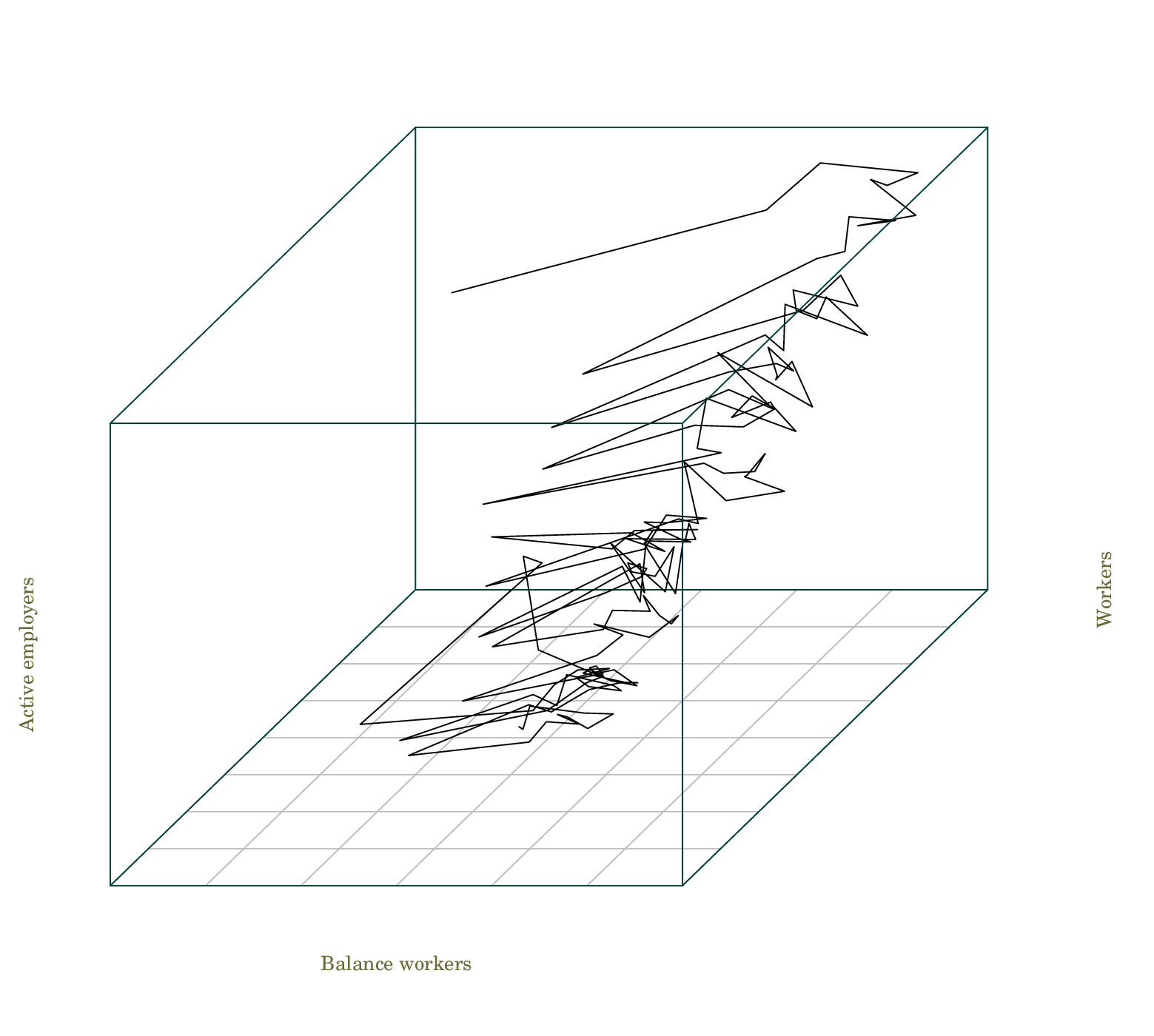}
\\
\small {Figure 1-a: {\it the relationship of workers, active employers and balance of workers.}} 
\end{figure}
\begin{center}
\footnotesize {{\bf Table 1: aggregate data of the CAGED database.}}
\begin{tabular}{rrrr}
Period & Balance workers & Workers & Active employers \\
01/1996 & -12.626 & 23.743.110 & 336.946 \\
02/1996 & -4.094 & 23.739.016 & 334.349 \\
03/1996 & 10.003 & 23.749.019 & 359.248 \\
04/1996 & 118.918 & 23.867.937 & 349.273 \\
05/1996 & 172.930 & 24.040.867 & 344.896 \\
06/1996 & 115.028 & 24.155.895 & 326.832 \\
07/1996 & 68.920 & 24.224.815 & 338.098 \\
08/1996 & 46.937 & 24.271.752 & 339.688 \\
09/1996 & 88.964 & 24.360.716 & 327.525 \\
10/1996 & 19.466 & 24.380.182 & 329.672 \\
11/1996 & -15.899 & 24.364.283 & 308.009 \\
12/1996 & -258.516 & 24.105.767 & 298.644 \\
\end{tabular} 
\end{center} 
The original Lotka-Volterra predator-prey model makes several simplifying assumptions; it was adapted in many ways, as we may see in the vast literature on predator-prey interactions from which the authors selected some of the more important publications (see references). In many cases the former assumptions are relaxed or rearranged to fit some particular dynamics. There is no doubt that the predator-prey model adapts, at least conceptually, the work-employment scenario, as it is very reasonable to admit that employers and workers are, after all, two populations conflicting with one another. Since the formulations of Karl Marx - with the analysis of the conflicts between worker and employer, and of the entailments of such conflicts with structural elements formed by the connections among politic, social and economic plains - we understand that conflicts belong to the dialectic nature of the capitalist work relations. Also Richard Goodwin, in a different approach, found in Marx arguments conceptual similarities that lead to his predator-prey dynamic model \cite{Goodwin}. 
\section{Methods and simulations}
\subsection{Formal representation}
The pertinent information such a model can render is whether or not population abundances tend to an equilibrium at which both, workers and employers, will coexist and survive with few conflicts. The adopted model was based on the approach of Blasius and Stone \cite{Blasius} about oscillations with Uniform Phase Evolution and Chaotic Amplitudes (UPCA). An ecological UPCA model was chosen because it is suitable for treatment of two correlated periodic phenomena with chaotic amplitude varying in time, however, almost constant amplitude-independent frequencies as hinted in Figure 1. We applied the model to investigate the complex temporal phase synchronization in work-employment system at Brazil. In other words, our approach was based on the theory of synchronization of chaotic oscillations, defined in the most general case \cite{Rosenblum} as the entanglement between the phases of two coupled systems, while the amplitudes stay over chaotic regime in time.
The real entanglement of phases in predator-prey relationships is much more complex than 
one could suppose at first. As pointed out by Holling, predator-prey interactions have shown that, in some cases of population densities, predators may control the number of preys, but this is not true to all density cases \cite{Holling}. However in Figure 1 populations grow and fluctuate in a net periodic way, the tests on the tangle between these populations goes far beyond the simplicity of that plot, as the reader can see in Figure 1-a.
The balance of workers is a finite difference that indicates whether the
work-employment scenario is favorable or not to the employee, and to what degree; it reflects the potential of the
Mean (work market) to sustain the system evolving, by analogy with the available natural resources in an ecological niche. In other words, if the balance is negative we believe that there is a retraction of work market to support occupied posts and, virtually, active employers. It is a variable that provides additional qualitative (positive or negative sign) information. Figure 1-a shows the active employers-workers-balance of workers relationship, illustrating the potential oscillations in these populations.
\begin{figure} [h]
\includegraphics[scale=0.52]{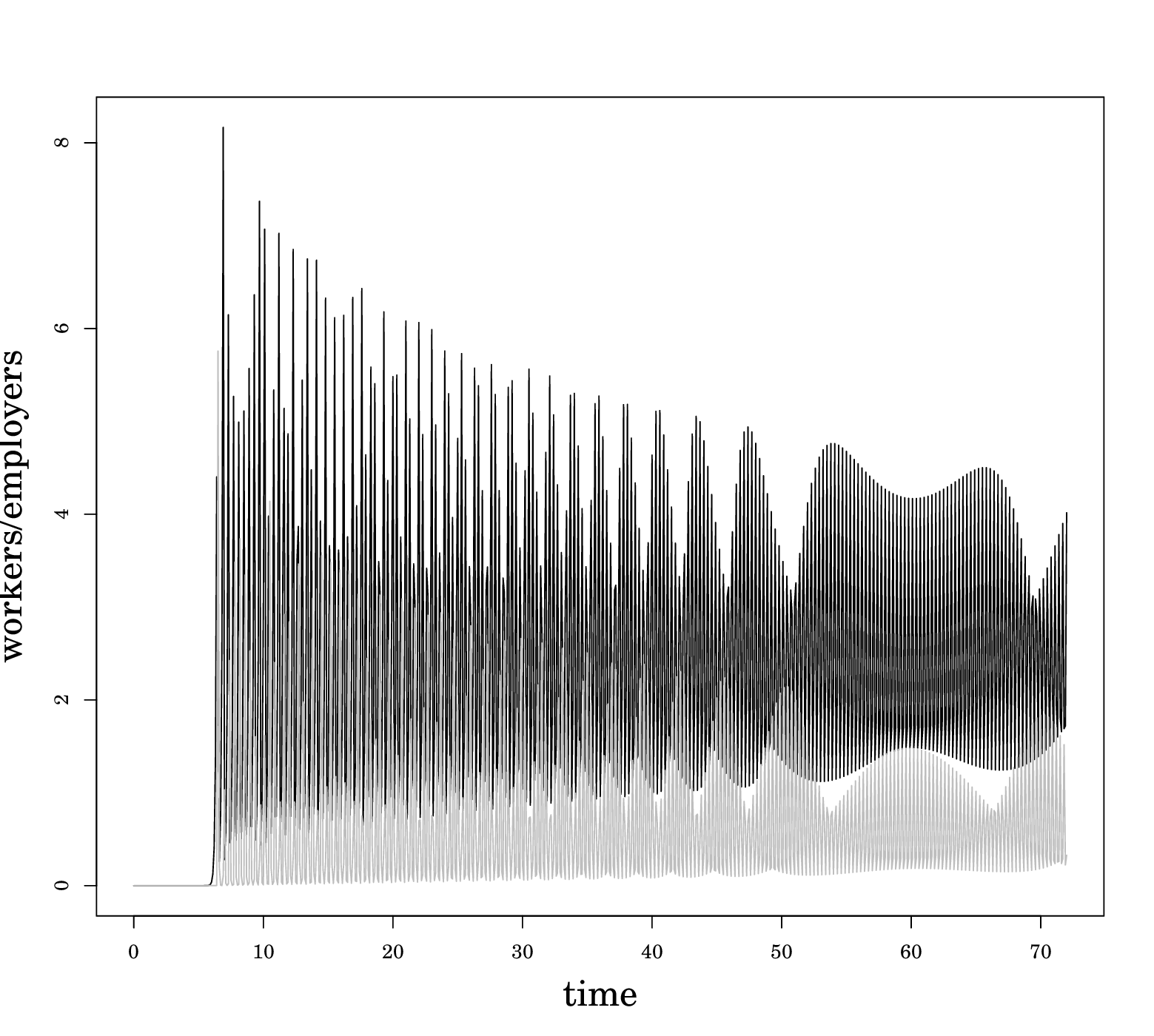}
\\
\small {Figure 2: {\it UPCA oscillations of employer (grey) and worker (black) dissimilarities with initial time-delayed growth ($\alpha _2=1.0$)}.} 
\end{figure}
\begin{figure} [h]
\includegraphics[scale=0.52]{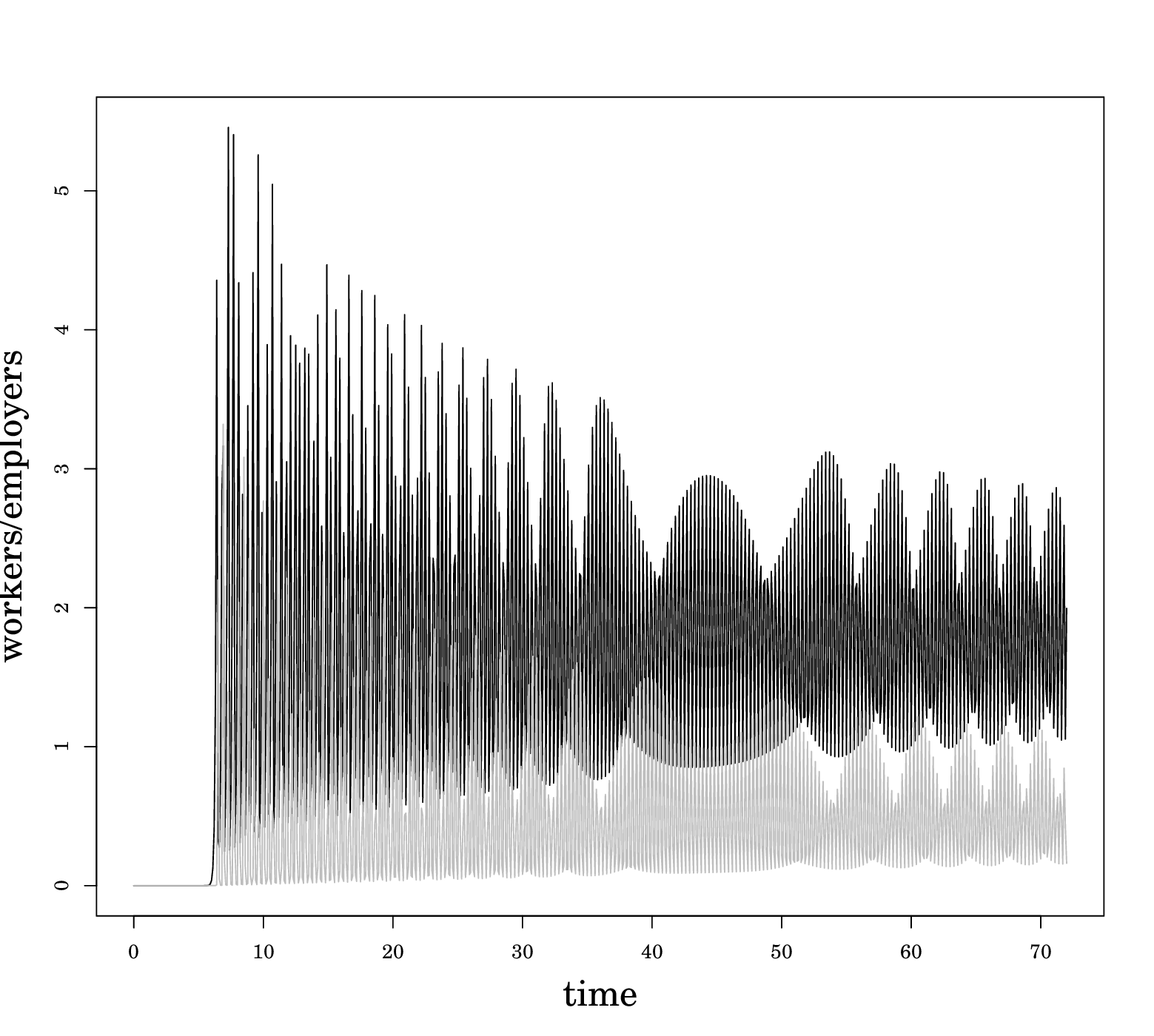}
\\
\small {Figure 3: {\it UPCA oscillations of employer (grey) and worker (black) dissimilarities with initial time-delayed growth ($\alpha _2=1.4$)}.} 
\end{figure}
\begin{figure} [h]
\includegraphics[scale=0.52]{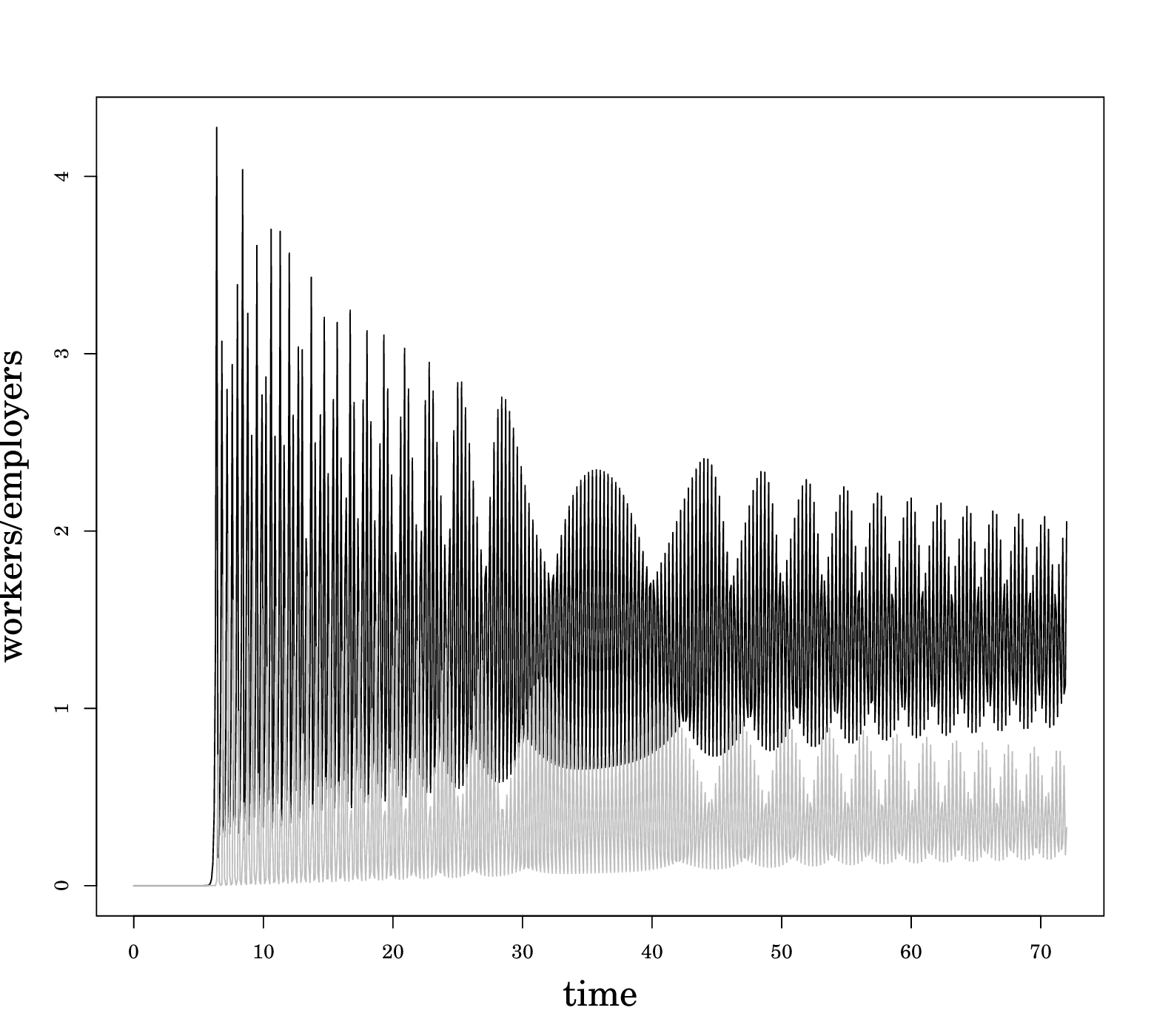}
\\
\small {Figure 4: {\it UPCA oscillations of employer (grey) and worker (black) dissimilarities with initial time-delayed growth ($\alpha _2=1.8$)}.} 
\end{figure}
\begin{figure} [h]
\includegraphics[scale=0.52]{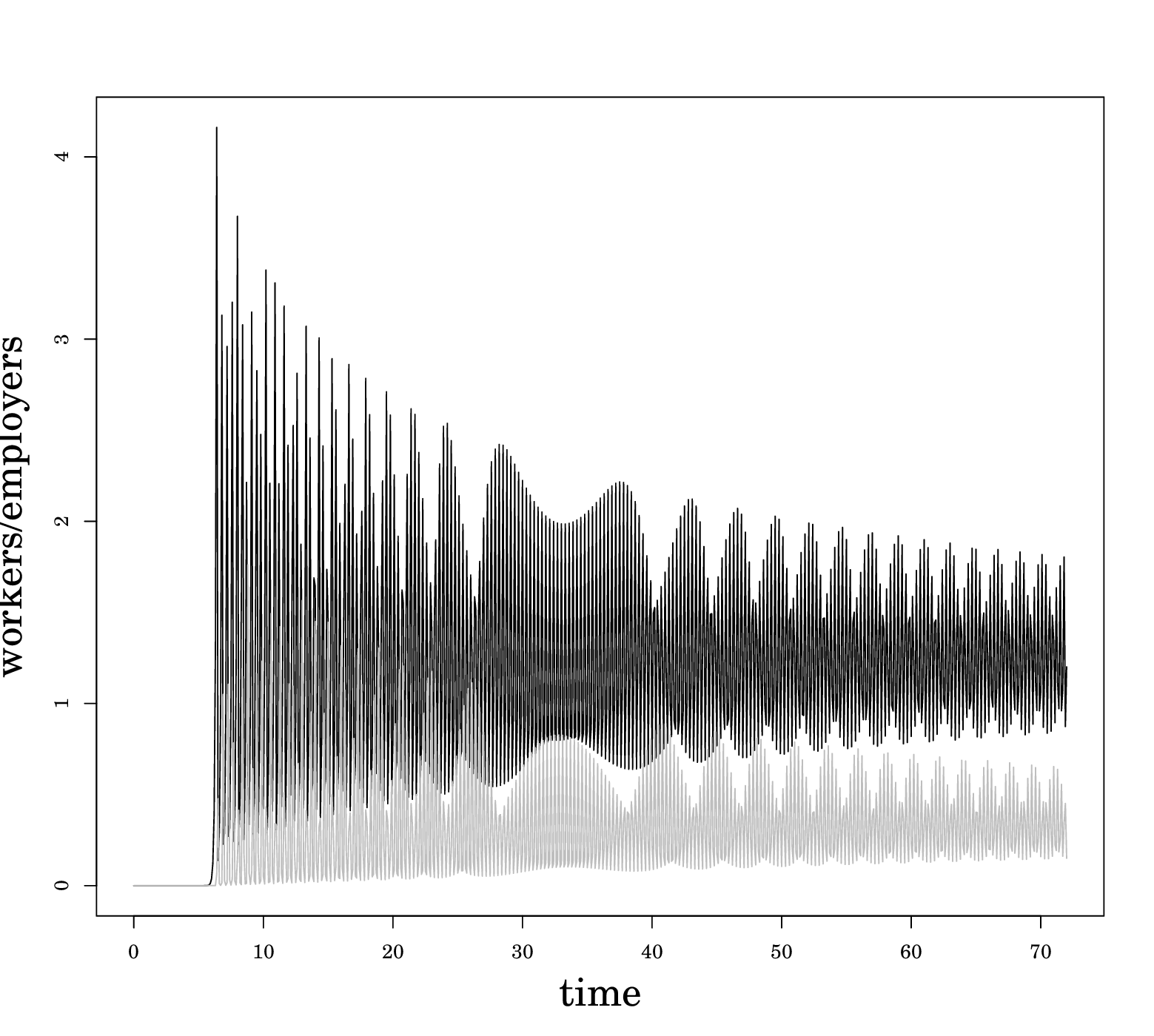}
\\
\small {Figure 5: {\it UPCA oscillations of employer (grey) and worker (black) dissimilarities with initial time-delayed growth ($\alpha _2=2.0$)}.} 
\end{figure}
\begin{figure} [h]
\includegraphics[scale=0.52]{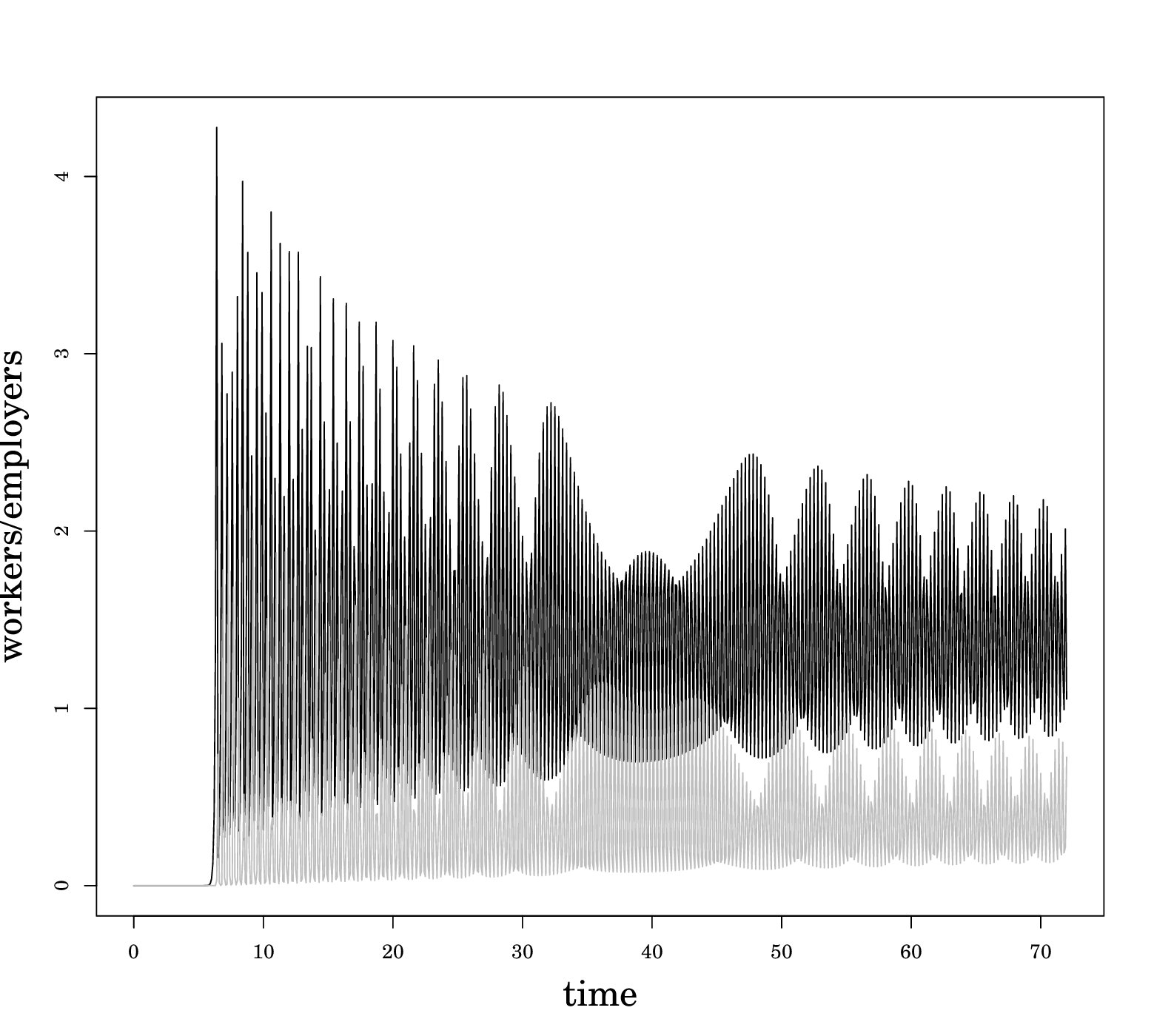}
\\
\small {Figure 5-a: {\it UPCA oscillations of employer (grey) and worker (black) dissimilarities with initial time-delayed growth and $k_2=0.00008$ ($\alpha _2=1.8$)}.} 
\end{figure}
The Lotka-Volterra-Serpa differential equations consider a state vector formed by the product between a geometric tracing vector and a dissimilarity vector of observables. We use pairwise dissimilarities between registers instead of the quantities directly stored in data base just to compute the relative differences which are really cyclic all along the time series. That is the basic difference from the classical Lotka-Volterra approach and the name was given by fellow researchers in honor of the author, Nilo Sylvio Costa Serpa. The dissimilarity matrix was made from the monthly populations of active employers and workers, and from the monthly balance of workers. This matrix computes all the pairwise dissimilarities (or "distances") between observations in the CAGED data set by means of the DAISY algorithm \cite{Kaufman}, that executes mixed measurements. We assume that the pairwise differences computed from the time registers determine the dynamical configuration of the system. Thereby, our model has three differential equations, one for the dissimilarity of balance of workers $u$ (resources), one for the dissimilarity of workers $v$ (preys) and one for the dissimilarity of active employers $w$ (predators), such as
\begin{equation}
\left\{ \begin{array}{l}
\dot u{\rm{ }} = {\rm{ }}a.u.u_o - {\rm{ }}\alpha _1 .{\rm{ }}f_1 (u.u_o ,{\rm{ }}v.v_o ,{\rm{ }}k_1 ) \\ 
\dot v{\rm{ }} = {\rm{ }} - b.v.v_o + {\rm{ }}\alpha _1 .{\rm{ }}f_1 (u.u_o ,{\rm{ }}v.v_o ,{\rm{ }}k_1 ){\rm{ }} - \\ 
- \alpha _2 .f_2 (v.v_o ,{\rm{ }}w.w_o ,{\rm{ }}k_2 ) \\ 
\dot w{\rm{ = }} - c.(w.w_o - {\rm{ }}w^\dag ){\rm{ }} + {\rm{ }}\alpha _2 .{\rm{ }}f_2 (v.v_o ,{\rm{ }}w.w_o ,{\rm{ }}k_2 ), \\ 
\end{array} \right.
\end{equation}
where $f_1$ and $f_2$ represent either the Lotka-Volterra term $f_i(x,y) = xy$ or the Holling type II term $f_i (x,y) = xy/(1 + k_i x)$, and $w^{\dag}$ denotes the minimum level of existent employers dissimilarity when there is scarcity of workpower or doldrums. The overdot indicates differentiation with respect to time. The quantities $a$, $b$ and $c$ are growth rates of balance of workers, workers, and employers respectively. The Holling type II functional response \cite{Holling} introduces a decelerating intake rate related to the assumption that the dissimilarities of active employers is limited by its capacity to regulate dissimilarities of workers. Comparing to the original equations applied by Blasius,
\[
\left\{ \begin{array}{l}
\dot u{\rm{ }} = {\rm{ }}a.u - {\rm{ }}\alpha _1 .{\rm{ }}f_1 (u,{\rm{ }}v,{\rm{ }}k_1 ) \\ 
\dot v{\rm{ }} = {\rm{ }} - b.v + {\rm{ }}\alpha _1 .{\rm{ }}f_1 (u,{\rm{ }}v,{\rm{ }}k_1 ){\rm{ }} - \alpha _2 .f_2 (v,{\rm{ }}w,{\rm{ }}k_2 ) \\ 
\dot w{\rm{ = }} - c.(w - {\rm{ }}w^\dag ){\rm{ }} + {\rm{ }}\alpha _2 .{\rm{ }}f_2 (v,{\rm{ }}w,{\rm{ }}k_2 ), \\ 
\end{array} \right.
\]
the reader can see that we merge variables $u, v, z$ with $u_0, v_0, z_0$ and this changes the final geometric form of the system integration according to de observed data. Figure 5-b shows an example of UPCA oscillations for the last system.
\begin{figure} [h]
\begin{center}
\includegraphics[scale=0.36]{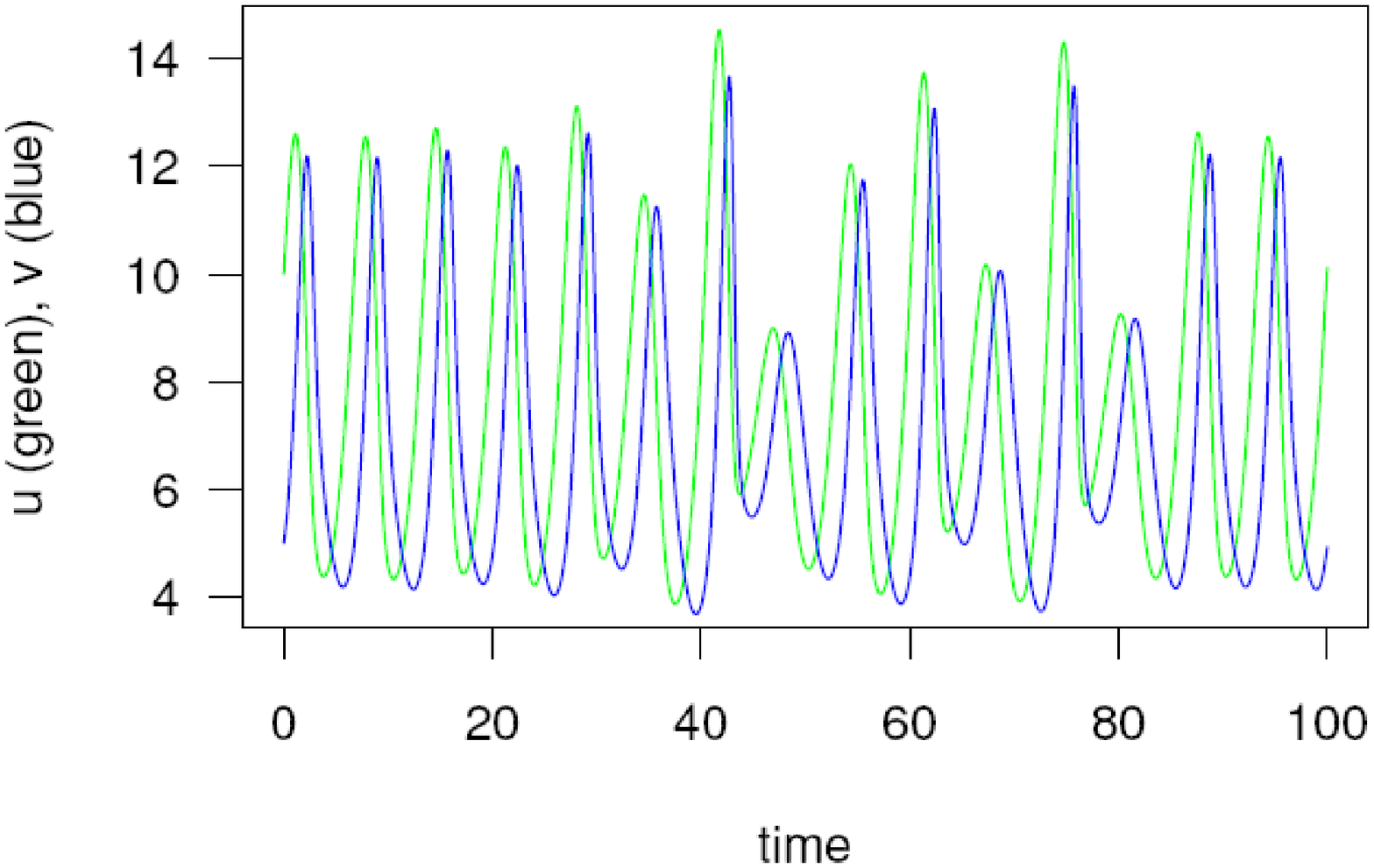}
\\
\small {Figure 5-a: {\it simple UPCA oscillations by Blasius {\it et al.} for variables $u$ and $v$.}} 
\end{center}
\end{figure}
In our model, the positive constant $k_2$ measure the carrying capacity of employers to absorb manpower; the positive constant $k_1$ measure the carrying capacity of the work market to accept new workers. As pointed out by Kenneth Arrow and colleagues \cite{Arrow},
\paragraph\
\paragraph\
\paragraph\
"{\it Carrying capacities in nature are not fixed, static, or simple relations. They are
contingent on technology, preferences, and the structure of production and consumption. They are also contingent on the everchanging state of interactions between the physical and biotic environment. A single number for human carrying capacity would be meaningless because the consequences of both human innovation and biological evolution
are inherently unknowable}".
\paragraph\
So, it is interesting to investigate what would happen for work-employment scenarios with very low carrying capacity of employers to create new posts. For instance, if we take $k_2=0$ we means that the employers are saturated or temporarily disabled to open new posts, although $k_1$ remains greater than zero, that is, the work market remains in flux, but mainly due to high turn over.
The contents of the differential variables were defined by means of the vectors, 
\begin{equation}
\begin{array}{l}
V\left( {u < - \omega [1],{\rm{ }}v < - \omega [2],{\rm{ }}w < - \omega [3]} \right), \\ 
U\left( {u_o < - \Pi [\delta ,1],{\rm{ }}v_o < - \Pi [\delta ,2],{\rm{ }}w_o < - \Pi [\delta ,3]} \right), \\ 
\end{array}
\end{equation}
being $\omega [j]$ ( $j$ in $[1, 2, 3]$) the geometric components describing the state trajectories and $\Pi [\delta ,2j]$ the $j$ dissimilarity components between observations with $\delta$ registers calculated according to the definition of Euclidian distance, that is,
\begin{equation}
\Pi [\delta ,j] \propto \left( {\sum\limits_{\kappa = 1}^d {\left| {x_{\delta \kappa } - x_{j\kappa } } \right|} ^r } \right)^{1/r} ,
\end{equation}
where $d$ is the phase space dimensionality; $x_{\delta\kappa}$ and $x_{j\kappa}$ are, respectively, the $\kappa-th$ components of the $\delta-th$ and $j-th$ registers. 
\subsection{The equilibrium condition}
No changes with respect to time in populations and resources, or in their dissimilarities, are represented by setting the equations equal to zero. So, the equilibrium condition for the system (1) requires,
\begin{equation}
\left\{ \begin{array}{l}
{\rm{ }}a.u.u_o - {\rm{ }}\alpha _1 .{\rm{ }}f_1 (u.u_o ,{\rm{ }}v.v_o ,{\rm{ }}k_1 ) = 0 \\ 
- b.v.v_o + {\rm{ }}\alpha _1 .{\rm{ }}f_1 (u.u_o ,{\rm{ }}v.v_o ,{\rm{ }}k_1 ){\rm{ }} - \\
- \alpha _2 .f_2 (v.v_o ,{\rm{ }}w.w_o ,{\rm{ }}k_2 ) = 0 \\ 
- c.(w.w_o - {\rm{ }}w^\dag ){\rm{ }} + {\rm{ }}\alpha _2 .{\rm{ }}f_2 (v.v_o ,{\rm{ }}w.w_o ,{\rm{ }}k_2 ) = 0 \\ 
\end{array} \right.
\end{equation}
Introducing the term of Lotka-Volterra (the simplest one), the equilibrium - where the dissimilarities of active employers, resources and workers do not vary - leads to,
\begin{equation}
\left\{ \begin{array}{l}
{\rm{ }}a.u.u_o - {\rm{ }}\alpha _1 .{\rm{ }}u_o .v_o .u.v = 0 \\ 
- b.v.v_o + {\rm{ }}\alpha _1 .u_o .v_o .u.v{\rm{ }} - \alpha _2 .v_o .w_o .v.w = 0 \\ 
- c.(w.w_o - {\rm{ }}w^\dag ){\rm{ }} + {\rm{ }}\alpha _2 .{\rm{ }}v_o .w_o .v.w = 0 \\ 
\end{array} \right.
\end{equation}
From the first equation we obtain, whereas population dissimilarity shall be never equal to zero, 
\begin{equation}
u.(a.u_o - {\rm{ }}\alpha _1 .{\rm{ }}u_o .v_o .v) = 0,
\end{equation}
\begin{equation}
a.u_o - {\rm{ }}\alpha _1 .{\rm{ }}u_o .v_o .v = 0,
\end{equation}
\begin{equation}
v = \frac{{a.{\rm{ }}u_o }}{{\alpha _1 .{\rm{ }}u_o .v_o }} = \frac{a}{{\alpha _1 .{\rm{ }}v_o }}.
\end{equation}
Taking the third equation we get,
\begin{equation}
- c.w.w_o + c.w^\dag + \alpha _2 .{\rm{ }}v_o .w_o .v.w = 0,
\end{equation}
\begin{equation}
w\left( { - c.w_o + \alpha _2 .{\rm{ }}v_o .w_o .v} \right) = - c.w^\dag , 
\end{equation}
\begin{equation}
w = \frac{{ - c.w^\dag }}{{ - \left( {c.w_o - \alpha _2 .{\rm{ }}v_o .w_o .v} \right)}} = \frac{{c.w^\dag }}{{c.w_o - \alpha _2 .{\rm{ }}v_o .w_o .v}}.
\end{equation}
Finally, taking the second equation it follows,
\begin{equation}
\alpha _1 .u_o .v_o .u.v{\rm{ = }}\alpha _2 .v_o .w_o .v.w + b.v.v_o, 
\end{equation}
\begin{equation}
u = \frac{{\alpha _2 .v_o .w_o .v.w + b.v.v_o }}{{\alpha _1 .u_o .v_o .v}},
\end{equation}
\begin{equation}
u = \frac{{\alpha _2 .w_o .w + b}}{{\alpha _1 .u_o }}.
\end{equation}
Substituting the value of $w$ given by equality (11), it comes,
\begin{equation}
u = \left( {\alpha _2 .w_o .\frac{{c.w^\dag }}{{c.w_o - \alpha _2 .{\rm{ }}v_o .w_o .v}} + b} \right)\frac{1}{{\alpha _1 .u_o }},
\end{equation}
\begin{equation}
u = \left( {\frac{{\alpha _2 .c.w^\dag }}{{c - \alpha _2 .{\rm{ }}v_o .v}} + b} \right)\frac{1}{{\alpha _1 .u_o }},
\end{equation}
\begin{equation}
u = \frac{{\alpha _2 .c.w^\dag + b.\left( {c - \alpha _2 .{\rm{ }}v_o .v} \right)}}{{\alpha _1 .u_o .\left( {c - \alpha _2 .{\rm{ }}v_o .v} \right)}}.
\end{equation}
Now, applying the result set by expression (8), we have,
\begin{equation}
u = \alpha _2 .c.w^\dag + b.\left( {\frac{{\alpha _1 .c - \alpha _2 .a}}{{\alpha _1 }}} \right)\frac{1}{{u_o .\left( {\alpha _1 .c - \alpha _2 .a} \right)}},
\end{equation}
\begin{equation}
u = \frac{{\alpha _1 .\alpha _2 .c.w^\dag + b.\left( {\alpha _1 .c - \alpha _2 .a} \right)}}{{\alpha _1 .u_o .\left( {\alpha _1 .c - \alpha _2 .a} \right)}}.
\end{equation} \\ \\
For the term of Holling type II, that is,
\begin{equation}
\begin{array}{l}
f_1 = u.u_0 .v.v_0 /\left( {1 + k_1 .u.u_0 } \right), \\ 
f_2 = v.v_0 .w.w_0 /\left( {1 + k_2 .v.v_0 } \right), \\ 
\end{array}
\end{equation}
the equilibrium sets,
\begin{equation}
\left\{ \begin{array}{l}
{\rm{ }}a.u.u_o - {\rm{ }}\alpha _1 .{\rm{ }}u.u_0 .v.v_0 /\left( {1 + k_1 .u.u_0 } \right) = 0 \\ 
- b.v.v_o + {\rm{ }}\alpha _1 .u.u_0 .v.v_0 /\left( {1 + k_1 .u.u_0 } \right) - \\
- \alpha _2 .v.v_0 .w.w_0 /\left( {1 + k_2 .v.v_0 } \right) = 0 \\ 
- c.(w.w_o - {\rm{ }}w^\dag ){\rm{ }} + {\rm{ }}\alpha _2 .{\rm{ }}v.v_0 .w.w_0 /\left( {1\left. { + k_2 .v.v_0 } \right) = 0} \right. \\ 
\end{array} \right.
\end{equation}
The system becomes much more complex. Nevertheless, we fix $k_2=0$, meaning that the employers are temporarily disabled to open new posts, which determines the solution as it follows. The second equation leads to,
\begin{equation}
w = \frac{{\alpha _1 .u.u_0 /\left( {1 + k_1 .u.u_0 } \right) - b}}{{\alpha _2 .w_0 }}. 
\end{equation}
The first equation gives,
\begin{equation}
a = \alpha _1 .v.v_0 /\left( {1 + k_1 .u.u_0 } \right),
\end{equation}
\begin{equation}
a.\left( {1 + k_1 .u.u_0 } \right) = \alpha _1 .v.v_0 , 
\end{equation}
\begin{equation}
1 + k_1 .u.u_0 = \frac{{\alpha _1 .v.v_0 }}{a} ,
\end{equation}
\begin{equation}
u = \frac{{\alpha _1 .v.v_0 - a}}{{u_0 .a.k_1 }} .
\end{equation}
Finally, the third equation sets $v$ as,
\begin{equation}
\alpha _2 .{\rm{ }}v.v_0 .w.w_0 /\left( {1\left. { + k_2 .v.v_0 } \right) = c.(w.w_o - {\rm{ }}w^\dag ),} \right.
\end{equation}
\begin{equation}
v = \frac{{c.(w.w_o - {\rm{ }}w^\dag )}}{{\alpha _2 .v_0 .w.w_0 }}.
\end{equation}
\begin{figure} [h]
\begin{center}
\includegraphics[scale=0.32]{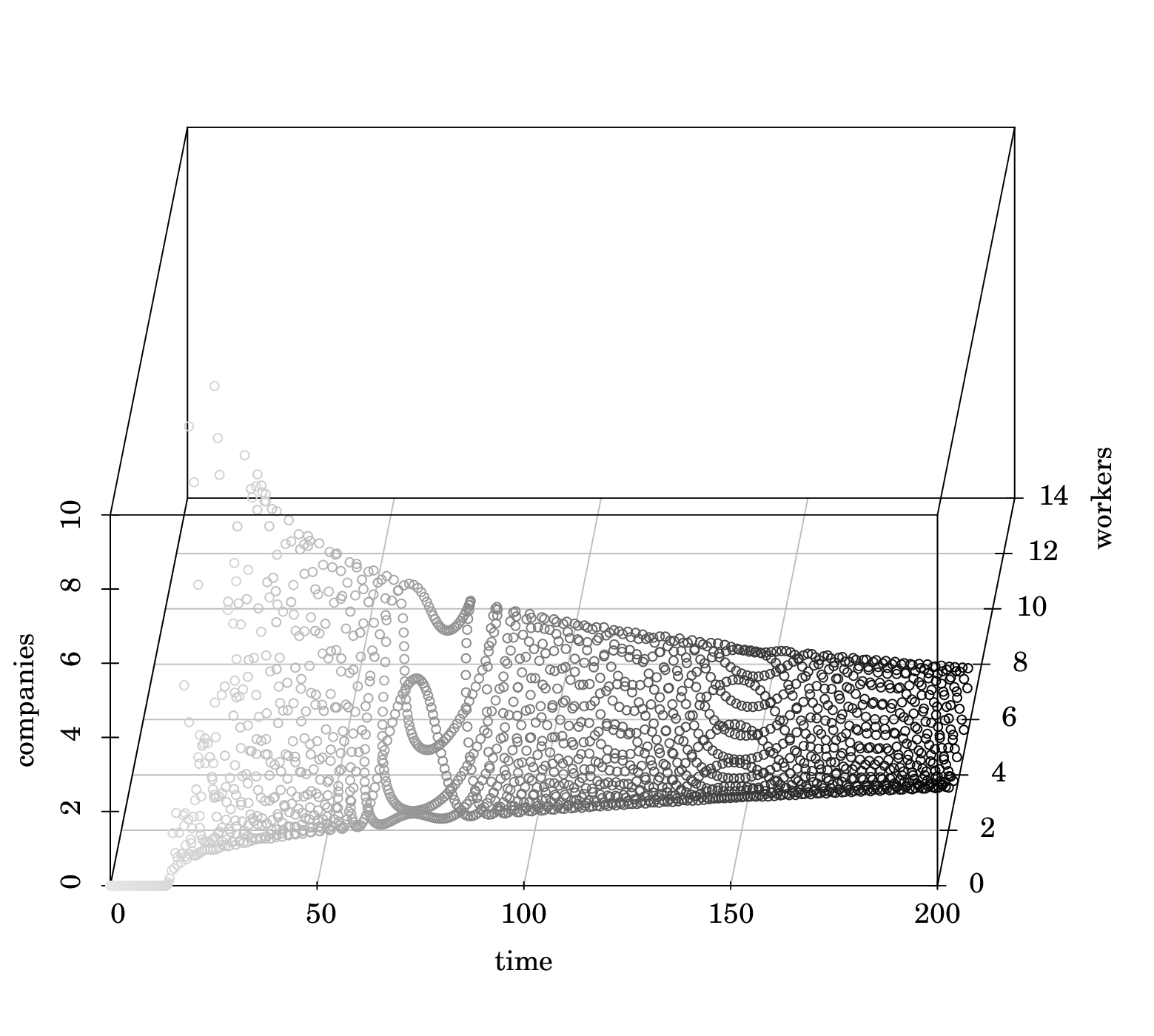}
\\
\small {Figure 6: {\it the system in three-dimensional phase space.}}
\end{center} 
\end{figure}
\begin{figure} [h]
\begin{center}
\includegraphics[scale=0.32]{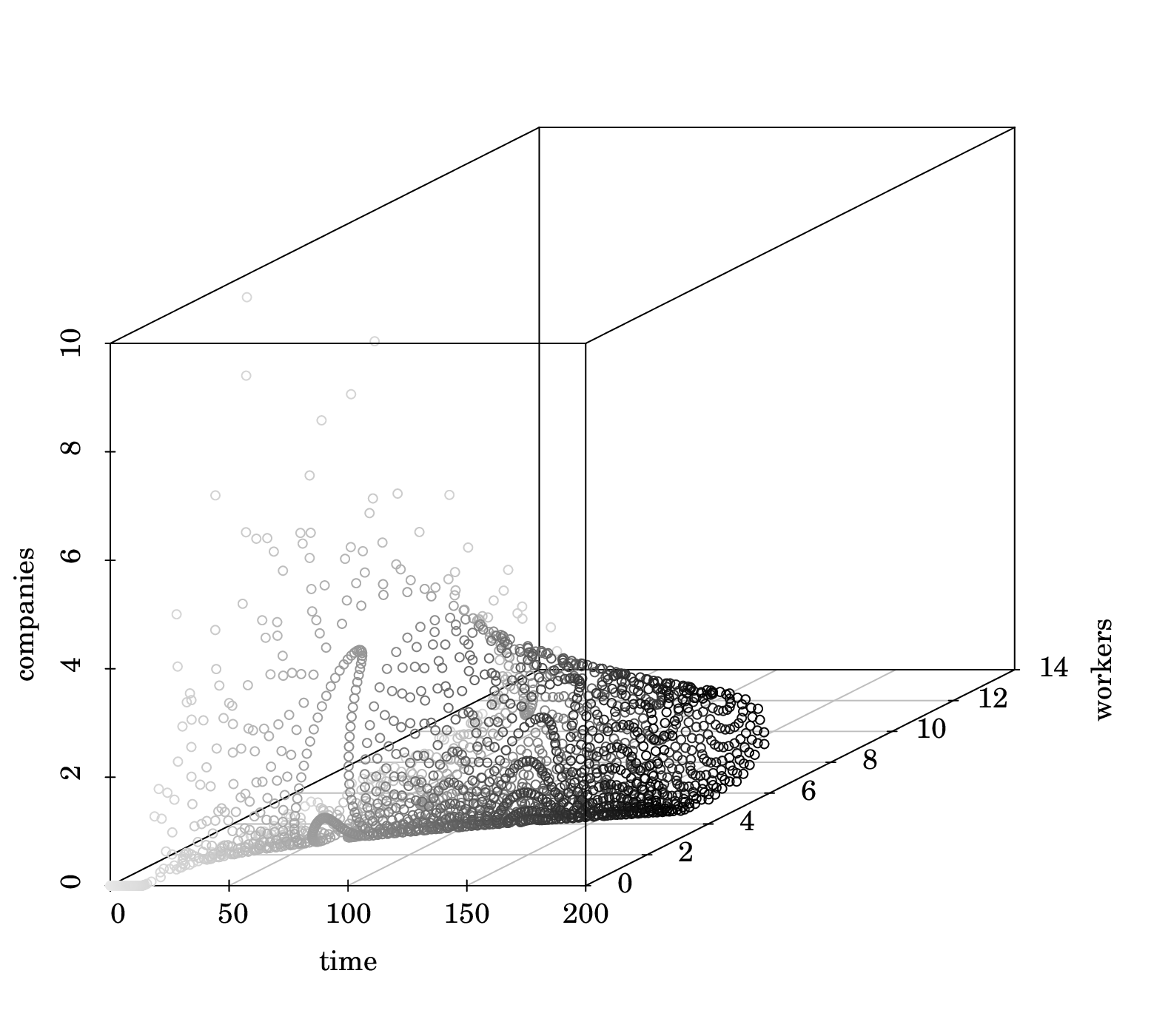}
\\
\small {Figure 7: {\it another vision of the same system.}} 
\end{center}
\end{figure}
\begin{figure} [h]
\begin{center}
\includegraphics[scale=0.32]{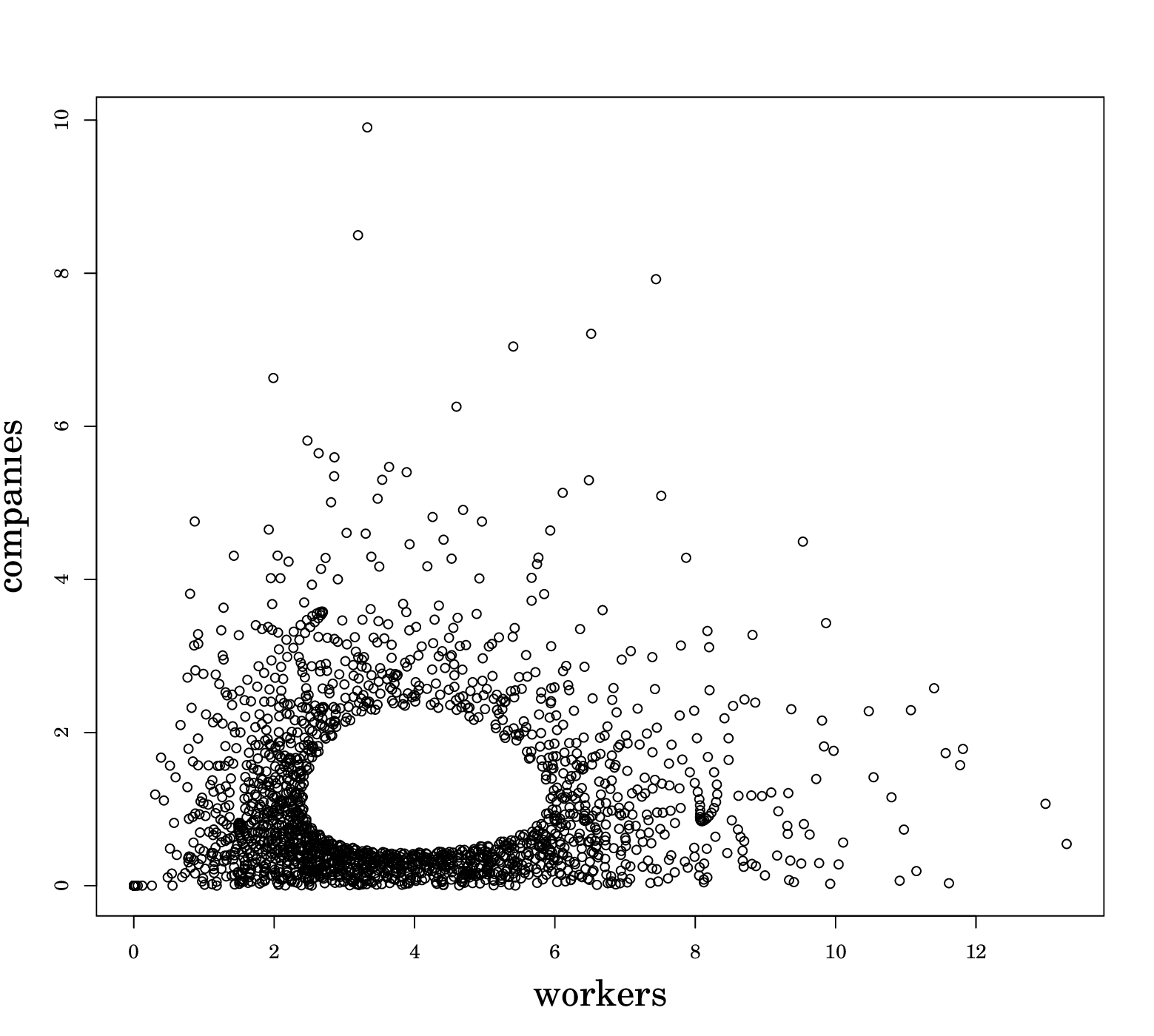}
\\
\small {Figure 8: {\it workers and employers in phase space with an attracting central curve.}}
\end{center} 
\end{figure}
\section{Discussion}
Insecurity generated by the employers and the work market, over many years of restructuring and adjustment to globalization as the Nineties, led to an exacerbation of individualism. Competition became more dominant than cooperation. Fear and instability certainly promoted higher levels of depression, while millions of people reached the productive age, pressing more and more the formal work market. Moreover, due to the neo-liberal offensive (1980-1990) to the resumption of the "Theory of Human Capital", backed by Gary Becker, the wage differences came to be seen as responsibility of the worker himself, who thus became the chief culprit for all economic ills. The spreading of innovations due to technological changes (the convergence of information technology and telecommunications and the continuing cost reductions) greatly affected the rate of hiring as a function of the resizing of the human effectives face to an accelerated automation. Also, Security in all the countries that care about their retirees do not support the significant increase in life expectancy, discouraging retirements and thus retaining jobs occupied for much longer. All these factors contributed to a more chaotic work-employment scenario in Brazil from mid-1990 until about 2002 and they corroborate the interpretation of the plots.
Many simulations are performed with vector $V$ describing the geometry of the dynamics (theoretical), intrinsic to the nature of the system itself, and vector $U$ storaging the measurable content (observational) of the dynamics. For the more realistic Holling type II functional response, with $k_2=0$, figures 2, 3, 4 and 5 show graphics of the general propensity of the system to less chaotic states accordingly the values $1.0$, $1.4$, $1.8$ and $2.0$ of the population interaction parameter $\alpha _2$. Figure 5-a shows similar patterns for $k_2=0.00008$. Note that in all these figures there is a transition state to the phase of greater relaxation of the system. The great phase transition occurs close to the middle of the time series, from which the system goes on a relaxation stage. Comparing figures 1 and 5, the best fit realy seems to be for $\alpha _2=2.0$ with phase transition around the zone between $50$ and $100$ months in figure 1. Figures 6, 7, and 8 show the system's phase space for $\alpha _2=1.0$. The exact fitting of this parameter is a matter of more studies, since predator-prey stability could come from many factors at play, such as migration of manpower, employers self-limiting their staffs, seasonal recruitment due to foreseen peaks of sales, and so on, in the same way as pointed out by Andrzej Pekalski \cite{Pekalski}in biological systems.
\section{Conclusion}
This study identified a predator-prey pattern in the work-employment system. The central equations applied are modified versions of those studied by Blasius {\it et al}, introducing pairwise dissimilarities between observations as a way to identify cyclic behaviors in the system. For carrying capacity $k_2=0$, simulations showed trajectories with initial chaotic behavior and tendency to final states of greater equilibrium, which means that, even with zero carrying capacity of employers, the dynamics of the work-employment system in Brazil has demonstrated, during the past 12-15 years, propensity to adjusting to less chaotic regimes. In part, this is due to the great migration in recent years to solidary economy and autonomous work. Present article analyzed the system dynamics under the interpretation of the population interaction parameter $\alpha _2$ as an index of the convergence between the aims of employers and workers or as an index of the equilibrium created by lower demand for formal employment. Of course, interactions and proliferation of workers and employers are not easily controlled, but the captivation of manpower is a variable that may be monitored and partially governed by public policies. Moreover, as Grafton and Echenique \cite{Echenique} pointed out, we have the problem to choose the more appropriate ecological model to describe the situation in study. In our opinion, as more is learned about the system, we must change adaptively the model in order to lower uncertainty. Far from exhausting the subject, it is clear that in its general form the model represents reasonably well the dynamics of predator-prey for the work-employment system in Brazil. We hope this work will serve as inspiration for further studies in order to compare the application of the model in different countries.
\section{Acknowledgements}
The authors acknowledge Dr. Sergio Alves Cotia for their comments on the manuscript and for his integral support to this work.
\begin{center}
$$
$$
$\diamondsuit\diamondsuit\diamondsuit$
\end{center} 

\end{document}